\documentclass    [twocolumn]{aastex61}
\usepackage{epsfig}

\begin{document}
\title{Cannonball model diagnosis of the short gamma ray burst 170817A}

\author{Shlomo Dado}
\affiliation{Physics Department, Technion, Haifa 32000, Israel}
\author{Arnon Dar}
\affiliation{Physics Department, Technion, Haifa 32000, Israel}
\author{A. De R\'ujula}
\affiliation{Theory Division, CERN, Geneva, Switzerland;
IFT, Universidad Aut\'onoma, Madrid, Spain}

\begin{abstract}

The rich and complex data obtained from multi-wavelength 
observations of SHB170817A, the short hard gamma ray burst (SHB) 
associated with GW170817 --the first neutron stars merger 
event detected in gravitational waves (GWs)-- are analyzed in the 
framework of the cannonball model of SHBs.  In this model a highly 
relativistic jet is launched by fall back matter on the nascent  neutron star 
(or black hole) into a surrounding glory (light from the surrounding 
wind nebula of the binary neutron stars) which was 
present already before the merger. The SHB was produced by inverse 
Compton scattering of  glory photons by the jet, which was viewed 
far off-axis. The fading glory, which produced the initial UVOIR 
afterglow, was powered by a neutron star remnant.
It was overtaken  by a late time X-ray, UVOIR and radio 
afterglow  produced by synchrotron radiation from the decelerating 
jet in the interstellar medium of the host galaxy. If the radio 
afterglow of SHB170817A was indeed produced by the jet, it should 
display a superluminal motion relative to the SHB location, still 
detectable in VLA and VLBI radio observations.

\end{abstract}


\section{Introduction} 
Gamma ray bursts (GRBs) were discovered fifty years ago by the
American spy satellites Vela (Klebesadel, Strong \& Olson 1973). For 
decades their origin and production mechanism remained  mysterious. 
Thirty years ago, Goodman, Dar and Nussinov (1987) suggested that 
GRBs may be produced in extragalactic neutron star 
mergers (NSMs) by an $e^+e^-\gamma$ fireball (Goodman 1986) formed by 
neutrino-antineutrino annihilation around the nascent compact object 
--a massive neutron star or a black hole. But shortly after the launch 
of the Compton Gamma-Ray Burst Observatory (CGRO), it became clear that 
such neutrino-annihilation fireballs are not powerful enough to produce 
observable GRBs at the very large cosmological distances indicated 
by the CGRO observations (Meegan et al.~1992). 

Consequently, Meszaros and Rees (1992) suggested that the $e^+e^-\gamma$ 
fireball produced in compact mergers may be collimated into a conical 
fireball by funneling through surrounding matter. Shaviv and Dar (1995), 
however, argued that GRBs are produced by inverse Compton scattering (ICS) 
of the surrounding light (glory) by narrowly collimated jets of highly 
relativistic plasmoids (cannonballs, CBs) of ordinary matter, launched in 
mergers of compact stars due to the emission of gravitational waves (GWs), 
in phase transition of neutron stars to quark stars in compact binaries 
following mass accretion, or in stripped-envelope core-collapse supernova 
explosions (Dar et al. 1992). Li and Paczynski (1998), countered that all 
GRBs are produced by macronovae, i.e., are the thermal radiation emitted 
from fireballs formed by the radioactive decay of r-process elements 
synthesized from the tidally-disrupted neutron-star surface material in 
compact binaries undergoing a merger by GW emission.

It has also been observed long ago that GRBs may be roughly classified into 
two distinct species, long-duration soft gamma-ray bursts (GRBs) that 
usually last more than 2 seconds and short hard bursts (SHBs) typically 
lasting less than 2 seconds (Norris et al.~1984; Kouveliotou et al.~1993), 
and that a large fraction of GRBs are produced in broad line stripped 
envelope supernova (SN) explosions of type Ic akin to SN1998bw (see Galama 
et al. 1998 for the first observed GRB-SN association; Dado et al.~2002, Zeh 
et al.~2004 and references therein for early photometric evidence; Stanek et 
al.~2003 and Hjorth et al.~2003 for the first spectroscopic evidence, and 
Della Valle et al.~2016 for a recent review. See also Dado, Dar \& De 
R\'ujula~2003 for the prediction of the discovery date and properties of the 
SN associated with GRB030329).

Based on indirect evidence, it was widely believed that SHBs were 
associated with NSMs (see, e.g., Fong and Berger 2013, Berger 
2014). In particular, a faint infrared emission from SHB130603B 
was claimed to be the first observational evidence for a 
macronova produced by a NSM (e.g.~Berger et al.~2013, Tanvir et 
al.~2013). The first indisputable NSM-SHB association, i.e., 
GW170817-SHB170817A, was observed only recently (von Kienlin et 
al.~2017, Abbott et al.~2017a,b,c, Goldstein et al.~2017). Two 
days before these ground breaking observations, Dado and Dar 
predicted (2017) that because of the relatively small horizon of 
Ligo-Virgo for detection of NSMs by gravitational waves, 
{\it only far off-axis SHBs or orphan afterglows, but not ordinary
SHBs, will accompany Ligo-Virgo detections of NSMs}.

The underlying process in the cannonball (CB) model of GRBs and 
SHBs is the ejection of highly-relativistic narrowly-collimated 
jets of CBs in stripped-envelope SNeIc and NSMs, respectively 
(e.g.,~Dar \& De Rujula~2004; Dado, Dar \& De R\'ujula~2002, 2009a; 
Dado, Dar \& De R\'ujula~2009b). The ejections of CBs may 
take place after the merger by accretion of fall back matter 
from mass ejections sometime before the end of the merger process. Such a delayed
ejection may explain the $\sim$1.74 s delay in the arrival of 
the electromagnetic radiation after the  arrival 
of the  gravitational waves (von Kienlin et al.~2017;
Abbott et al.~2017a,b,c; Goldstein et al.~2017). 

In the CB model, the gamma-ray generating mechanism is ICS of a glory light 
(light scattered or emitted by winds or earlier mass ejections). The natural 
candidate producing such a glory is a pulsar wind nebula (Weiler \& Panagia 
1978, for a review, see, e.g., Gaensler \& Slane 2006). Such a pulsar wind 
nebula (PWN) absorbs the magnetic dipole radiation and relativistic wind, 
which are emitted from the pulsar and converts most of its spin-down energy 
to synchrotron radiation in the radio, optical and X-ray bands, with 
approximately a broken power-law spectrum or an exponentially cutoff 
power-law spectrum and peak energy flux around 1 eV (see, e.g., 
Macias-Perez et al. 2010; Tanaka \& Takahara 2010 and references therein).

The observed duration of the SHB pulses in ordinary SHBs requires 
a glory of a typical size $\sim\!10^{15}$ cm. Such a relatively 
small size PWN may be quite natural for very compact neutron stars  
(n*s) binaries.  Assuming that both neutron stars may be approximated 
as point masses, circular n*n* binary orbit decays at a 
rate $da/dt\!=\!-a/t_{GW}$, where $a$ is the binary separation 
and the merger timescale due to the gravitational radiation is 
given in geometrized units (G=c=1) by $t_{GW}\!=\!(5/64)a^4/\mu\, 
M^2$.  $M\!=\!M_1\!+\!M_2$ is the total mass and 
$\mu\!=\!M_1M_2/(M_1\!+\!M_2)$ is the reduced mass of the n*n* 
binary. For canonical neutron stars, with 
$M_1\!=\!M_2\!=\!1.4M_\odot$ and initial separation $a$, the 
merger time due to gravitational radiation is $t_m\!=\!\int\! 
(dt/da)da\!\approx\! 1.76\,(a/2R_\odot)^4$ Gy. This suggests that 
n*n* binaries with a merger time much shorter than the Hubble 
time must be born in very compact binaries. Such compact binaries 
may be formed either in a single SN explosion of a massive star 
by fission of its fast rotating core during its rapid collapse, 
or in two separate SN explosions in massive star binaries where 
dynamical friction in a common envelope phase shrinks the binary 
separation (e.g., Bhattacharya \& van den Heuvel 1991; Lorimer 
2008 and references therein).

In this paper we show that ICS of photons emitted by a PWN 
surrounding merging neutron stars, by a highly relativistic jet of CBs
launched in the NSM and viewed far off-axis, can explain the prompt 
emission in SHB170817A, provided that the typical size of the PWN 
around NSMs is $R\sim 10^{15}$ cm, the PWN  luminosity peaks around
1 eV, and the typical bulk motion Lorentz factor of the CBs, which produce 
SHBs, is  $\gamma\sim 10^3$ (Dado et al. 2009b). The PWN may have a 
disk-like shape, or a torus-like shape, or even a more 
complicated shape. 
  
In our CB model analysis, the absence of an extended X-ray emission 
in SHB170817A (Troja et al.~2017a,b), following the prompt 
$\gamma$-ray emission, is explained by the merger site {\it not} 
being in a densely illuminated region (Levan et al.~2017), unlike 
globular clusters where probably a considerable fraction of SHBs 
occur (Dado et al. 2009b).

The observed early time UVOIR afterglow is explained by light 
emission from an expanding fireball formed and powered by the ejecta,  
winds, radiations and high energy particles emitted from the merging 
n*s and the nascent n*, before and after the merger, respectively.

The observed late-time radio (Hallinan et al. 2017, Mooley et al. 2018)
to X-ray (Troja et al.~2017a,b) afterglow, is explained 
by synchrotron radiation from the far off-axis decelerating jet.
If the radio afterglow is emitted by the 
jet, it should display a superluminal motion relative to the SHB's 
location (Dar \& De R\'ujula 2000a, Dado, Dar \& De R\'ujula 2016) 
hopefully still detectable by VLA and VLBI radio observations.

\section{The prompt emission} 

In the CB model SHBs and GRBs share many 
properties, since they are produced by the same mechanism: ICS of 
ambient light by a narrow jet of CBs with large Lorentz factors 
$\gamma\!\gg\! 1$. Their most probable viewing angles relative to the 
approaching-jet direction are $\theta\!\approx\!1/\gamma$ and the 
polarization of their radiation is predicted to be linear and large: 
$\Pi\!=\!2\,\gamma^2\,\theta^2/(1\!+\!\gamma^4\,\theta^4)\!\approx\! 
1$ (Shaviv \& Dar~1995; Dar \& De R\'ujula~2004 and references 
therein), while very near axis and far off-axis GRBs and SHBs
should display a small polarization.

The CB model entails very simple correlations between the main 
observables of SHBs and GRBs (Dar \& De R\'ujula 2000a). For a 
burst at redshift $z$ and Doppler factor $\delta\simeq 2\,\gamma/ 
(1\!+\!\gamma^2\,\theta^2)$, for instance, the peak energy of their 
time-integrated energy spectrum satisfies 
$(1+z)\,E_p\!\propto\!\gamma\,\delta$, while their 
isotropic-equivalent total gamma-ray energy is 
$E_{iso}\!\propto\!\gamma\,\delta^3$. Hence, ordinary GRBs and 
SHBs, mostly viewed from an angle $\theta\!\approx\!1/\gamma$, obey
\begin{equation}
(1+z)\,E_p\!\propto\! [E_{iso}]^{1/2},
\label{Eq1}
\end{equation}
while the far off-axis ($\theta^2\gg 1/\gamma^2$) ones satisfy
\begin{equation}
(1+z)\,E_p\!\propto\! [E_{iso}]^{1/3}.
\label{Eq2}
\end{equation}
Updated results on the correlation of Eq.(1), 
later empirically discovered by Amati et al.~(2002), 
are shown in Figure 1. They satisfy well 
the CB model predicted correlation for ordinary GRBs.

In Figure 2 we plot the 
observations for the correlation Eq.(2)
for far off-axis GRBs (Dar \& De R\'ujula 2000a), 
which is also well satisfied.

\begin{figure}[]
\centering
\epsfig{file=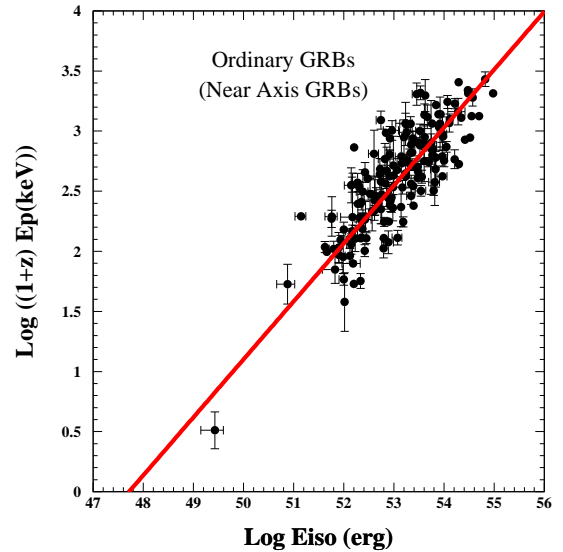,width=8.cm,height=8.cm}
\caption{The $[E_p,E_{iso}]$ correlation for 
ordinary GRBs  viewed near axis. The line is 
the CB model prediction of Eq.(1).}
\label{Fig1}
\end{figure}

\begin{figure}[]
\centering
\epsfig{file=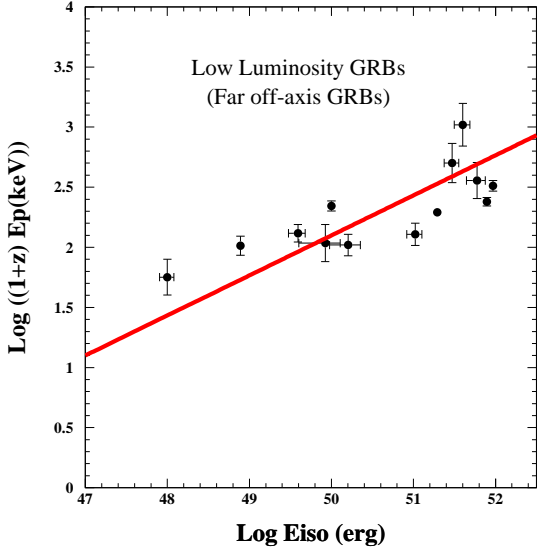,width=8.cm,height=8.cm}
\caption{The $[E_p,E_{iso}]$ correlation of low luminosity 
(far off-axis) GRBs. The line is the CB model prediction of 
Eq.(2).}
\label{fig2}
\end{figure}

In Figure 3 we plot results for the entire 
population of SHBs with known $z$, $E_p$ and $E_{iso}$, along with 
the predictions of Eqs.(1),(2). 
Ordinary (near axis) SHBs satisfy Eq.(1). The 
prediction of Eq.(2) for low luminosity (far 
off-axis) SHBs cannot be tested  because of the incompleteness 
of the current data (unknown $E_p$ and/or $z$) on the few low
luminosity SHBs.  

\begin{figure}[]
\centering
\epsfig{file=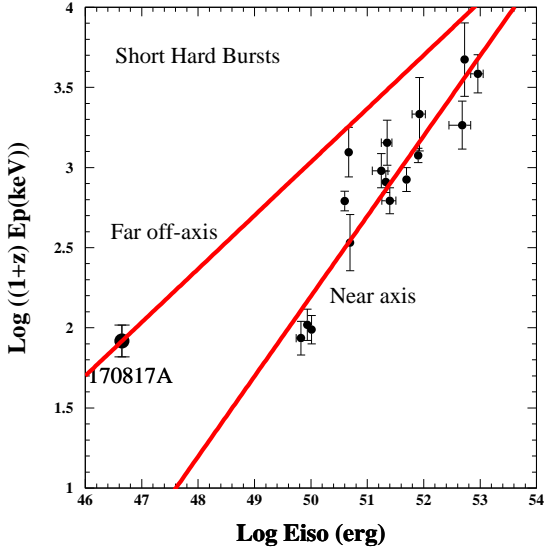,width=8.cm,height=8.cm}
\caption{The $[E_p,E_{iso}]$ correlation for SHBs with known redshift.
The lines are the CB model prediction of Eqs.(1),(2)
for near and far off axis cases.}
\label{fig3}
\end{figure}
Far off-axis viewing angles yield $\gamma\,\delta\!\approx\! 
2/\theta^2$. Hence, ICS of a standard candle glory light with a 
peak photon energy $\epsilon_p\!\approx\!1\, eV$, by a jet viewed 
from far off-axis yields an SHBs with $(1+z)E_p\!\approx\! 2\, 
eV/\theta^2$. Thus, the $T_{90}$ measured  $E_p\!=\!82\!\pm\!23$ keV
of SHB170817A (von Kienlin et al.~2017) implies a viewing angle 
$\theta\!=\! \sqrt{2/82\times 10^3}\!\approx\! 5\times 10^{-3}$ 
(the value $E_p\! \approx\! 65\!+\!35/\!-\!14$ keV, estimated by 
Pozanenko et al.~(2018) from the same data yields $\theta\!\approx\!
5.5 \times 10^{-3}$). 
Such $E_p$ is well above that 
expected from Eq.(1). But it is what is expected from Eq.(2), if 
the relatively small $E_{iso}\!\approx\! 5.4\times 10^{46}$ erg (Goldstein 
et al. 2017) is due to being viewed far off-axis. 
The same conclusion can also be drawn from its observed late-time  X-ray 
afterglow (Troja et al.~2017a,b), which is simply impossible to fit in the CB 
model with $\gamma\,\theta\!\sim\! 1$, as we shall see in detail in 
Section 6. Such a viewing angle is similar to that of the 
first-detected far off-axis GRB 980425. In many respects, 
SHB170817A-NSM170817 is similar to GRB980425-SN1998bw, both of 
which were the first of their kind.

In the Thomson regime ($2\gamma\epsilon\!\ll\!m_ec^2)$, the 
distribution of the incident photons after Compton scattering in 
the CB is nearly isotropic. In the observer frame this distribution 
becomes $dN_\gamma/d\Omega\!=\!N_\gamma\,\delta^2/4\,\pi\,.$ The 
canonical $\gamma\!=\!1000$ of ordinary SHBs (e.g., Dado et al. 
2009b) and the viewing angle $\theta\!=\!5$ mrad yield 
$\delta\!=\!80$. Consequently, in the CB model (Dar \& De R\'ujula 
2004), the measured $E_{iso}\!\approx\! 5.4\times 10^{46}$ erg 
(Goldstein et al.~2017) and $E_p\!=\!82$ keV (von Kienlin et 
al.~2017) in SHB170817A, yield a total number of ICS photons 
$N_\gamma\!=\!E_{iso}/E_p\,\delta^2\!\approx\!6.4\times 10^{49}$, 
and a total  $\gamma$-ray energy $E_\gamma\!\approx\! 
\epsilon_p\,\gamma^2\,N_\gamma\! \approx\! 1.0\times 10^{44}$ erg. 

Assuming that ejected CBs in NSMs are made of n* surface material, 
i.e., mainly iron nuclei with roughly equal number of protons and 
neutrons (Chamel \& Haensel 2008, and references therein), and that 
the kinetic energy of the CB electrons powers 
the prompt $\gamma$-ray emission by ICS of glory light, then, in the CB 
model, the estimated kinetic energy of the CB in SHB170817A was 
$E_k\!\approx\! 2\,m_p\,E_\gamma /m_e\!\approx\! 3.7\times 10^{47}$ 
erg, and its baryon number  was 
$N_b\!\approx\! E_k/m_p  c^2\gamma\!\approx\! 2.5\times 10^{47}$.

The peak time $\Delta$ of the prompt emission pulse is reached when 
the CB becomes transparent to photons, i.e., when the photon's mean 
diffusion time $t_d$ out of the CB satisfies 
$R_{cb}^2/(c\,\lambda)=\!\Delta $.
The photon's mean free path $\lambda$ that is dominated by Thomson
scattering on free electrons is given by 
$\lambda\!=\!4\,\pi\, R_{cb}^3 /(N_e\,\sigma_{_T})$
where $\sigma_{_T}\!=\!0.67\times 10^{-24}\, {\rm cm^2}$ 
is Thomson cross section, and the electron number of the CB satisfies 
$N_e\!\approx\!~N_b/2$.  Thus, 
the  peak time of the ICS pulse in SHB170817A
formed by  CB with  $N_e\!\approx \!N_b/2\approx 1.25\times 10^{47}$,
which is expanding  with a speed of sound in a relativistic  gas, 
$v\!=\!c/\sqrt{3}$, is expected to occur at
$\Delta\!\approx \!3\,\sqrt{3}\, N_e\, \sigma_{_T}\,/4\,\pi\, 
c^2\,\delta=0.69$ s. 

The mean  photon density $n_g$ of the glory in the volume
$V$ where from the SHB photons were scattered, 
satisfies $N_\gamma\!=\! n_g\, V\!\approx 
\!n_g\,\pi\,\gamma\,\delta^3\,c^3\,\Delta^3/9$.
It yields $n_g\!\approx\! 4.0\times 10^{10}\,{\rm cm^{-3}}$ for
a  peak time $\Delta\!\approx\!0.69$ s of the  prompt emission 
pulse of SHB170817A. 

Note that while the CB model can predict the approximate pulse 
shape of single pulses (given their $E_{iso}$, $E_p$ and $R$), it 
cannot predict the entire light curve of multi-pulse GRBs and SHBs, 
because neither the time sequence, nor the emission directions, nor 
the physical parameters of CBs are predictable. Moreover, a shotgun 
configuration of the emitted CBs, and/or a precession of the emission 
direction, as observed in pulsars and microquasars, combined with 
relativistic beaming, can make only a small fraction of the emitted 
CBs visible. That, and the very near distances of SHB170817A and 
GRB980425 can enhance the detection rate of such events by a large 
factor compared to that estimated from the assumption that only the 
CBs which produced the observed light curves were actually emitted.

For instance, long GRBs are detected by the Fermi GBM at a rate 
roughly 6 times larger than the detection rate of SHBs. The current 
horizon of Ligo-Virgo for detection of n*n* mergers is roughly 180 
Mpc. During the past 12 years, 2 GRBs were detected with a redshift 
within this horizon (060218 and 111005A). If GRBs and SHBs have the 
same redshift distribution, then an event like SHB170817A should be 
detected by a Fermi GBM like detectors at a rate roughly once in 30 
years, unless NSM170817 has launched many more CBs in a 
shotgun configuration or in succession along a precessing axis, out 
of which only 2 were detected by the Fermi GBM (von Kienlin et al. 
2017; Goldstein et al.~2017).

\section{Pulse shape}

In the absence of a detailed knowledge of the properties of the glory 
around n*n* mergers, and in view of the small statistics of the 
$\gamma$-ray data on SHB170817A (Goldstein et al. 2017), we perform only an 
approximate test of the CB model predictions for the temporal and spectral 
behaviors of  its prompt $\gamma$-ray emission.

The observed  pulse-shape produced by ICS of glory 
light with an exponentially cut off power law (CPL) spectrum, 
$dn_g/d\epsilon\!\propto\! \epsilon^{-\alpha}\,exp(-\epsilon/\epsilon_p)$ at 
redshift $z$, by a CB is given approximately (see, e.g., Eq.~(12) in Dado et al.~2009b) by 
\begin{equation}
E{d^2N_\gamma\over 
dE\,dt}\!\propto\!{t^2\over(t^2\!+\!\Delta^2)^2} \, E^{1-\alpha}\,exp(-E/E_p(t)) 
\label{Eq3} 
\end{equation} 
where $\Delta$ is approximately the peak time of the pulse in the observer 
frame, which occurs when the CB becomes transparent to its internal
radiation and $E_p\!\approx\! E_p(t\!=\!\Delta)$.

In Eq.(3), the early temporal rise like $t^2$ is produced by the increasing 
cross section, $\pi\, R_{CB}^2\!\propto\! t^2$, of the fast expanding CB when it 
is still opaque to radiation (mainly due to Compton scattering from its 
electrons when it is still very hot and completely ionized). When the CB 
becomes transparent to radiation due to its fast expansion, its cross sectios 
for ICS becomes $\sigma_{_T}\,N_e$ where $\sigma_{_T}$
is the Thomson cross section, and $N_e$,  the
electron number of the CB, is roughly equal to half its baryon number $N_b$. 
That, and the density of the ambient photons, which for a distance  
$r\!=\!\gamma\,\delta\,c\,t/(1\!+\!z)\!>\! R$ decreases like 
$n_g(r)\!\approx\!n_g(0)\,(R/r)^2\!\propto\! t^{-2}$, produce the temporal 
decline like $t^{-2}$.
 
The ICS of glory photons with energy $\epsilon$ by the CB electrons 
boosts their energies to observed energies, which satisfy
$E\!=\!\gamma\,\delta (1+\beta \cos\theta)\epsilon /(1+z)$. 
The unknown geometry of PWNs of 
n*n* binaries can be very complex and different in different SHBs.
The glory, however, may attain a more universal shape, in particular 
outside the PWNs. To illustrate the effect of the increasing anisotropy 
of the glory photons when a CB moves away from its launch point, 
consider for simplicity a CB launched along the axis of a  
torus-like PWN, as illustrated in Figure 4. 

\begin{figure}[]
\centering
\epsfig{file=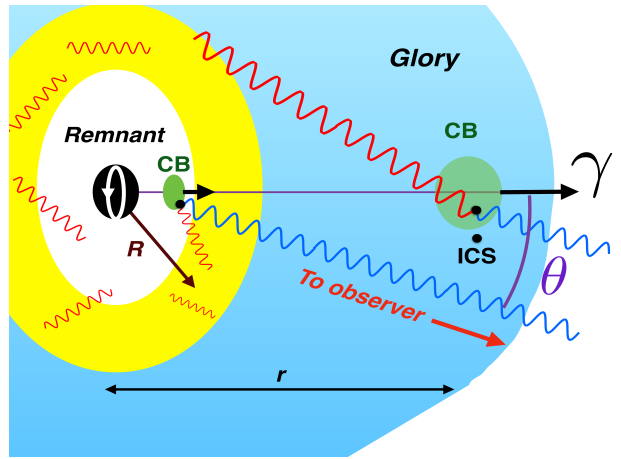,width=8.cm,height=6.cm}
\caption{A CB as it crosses and exits the
blue glory of a yellow toroidal PWN around NSM. The CB's 
electrons Compton up-scatter glory photons 
with incident angles which
decrease with increasing distance from the CB launch point.}
\label{Fig4}
\end{figure}

At the center of the PWN where $r\!=\!0$, 
the glory photons incident on the CB at an angle $\pi/2$ relative to 
its direction of motion.  At a distance $r\!>\! 0$
from the center of the PWN, 
the glory photons intercept the CB at angles that satisfy 
$\cos\theta \!=\!-\! r/\sqrt{r^2\!+\! R^2}$, which yields 
the t-dependence
\begin{equation}
E_p(t)\!=\!E_p(0)[1\!-\!t/\sqrt{t^2\!+\!\tau^2}]
\label{Eq4}
\end{equation} 
with  $\tau\!=\!R\, (1+z)/\gamma\,\delta\,c$.
Hence,  ICS of a glory with a CPL spectrum by a highly relativistic CB yields 
\begin{equation}
E{d^2N_\gamma\over dt\,dE}\!\propto\!{t^2\over(t^2\!+\!\Delta^2)^2}
 E^{1-\alpha}\exp(-E/E_p(0)[1\!-\!t/\sqrt{t^2\!+\!\tau^2}])
\label{Eq5}
\end{equation}
and $E_p\!\approx\! E_p(t\!=\!\Delta)$.
For $\alpha$ not very different from 1,
integration of $d^2N(E,t)/dE\,dt$ from $E\!=\!E_m$ upwards 
yields, 
\begin{equation}
N(t,E\!>\!E_m)\!\propto\!{t^2\over (t^2+\Delta^2)^2}\,exp(\!-\!E_m/E_p(t))
\label{Eq6}
\end{equation}
where $E_p(t)$ is given by Eq.(4).
\begin{figure}[]
\centering
\epsfig{file=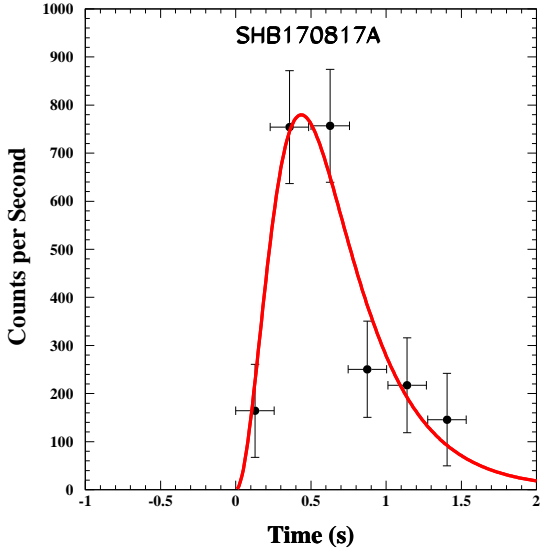,width=8.cm,height=8.cm}
\caption{Comparison of the pulse shape for $E_m\!=\!50$ keV
of the first pulse of SHB170817A measured by Goldstein et al.~(2017) 
and the CB model pulse shape as given by Eq.(6).}
\label{Fig5}
\end{figure}
A best fit of  Eq.(6) 
to the observed pulse shape (Goldstein et al.~2017) for $E_m\!=\!50$ keV,
which is shown in Figure 5, returns $\Delta\!=\!0.54$ s
$\tau\!=\!0.65$ s, and $E_p(0)\!=\!260$ keV.
The best fit value of $\Delta$ yields 
$E_p\!\approx\! E_p(\Delta)\!=\!94$ keV,
while $\tau\!=\!0.65$ s yields $R\!\approx\!1.6\times 10^{15}$ cm.

\section{No extended emission}
A considerable fraction of SHBs show an extended emission (EE)  
after the prompt SHB (Villasenor et al. 2005; Norris \& Bonnell 2006;
Fong \&  E. Berger 2013; Berger 2014  for a review).
Such SHBs may take place in rich star clusters or
globular clusters (GCs) (Dado, Dar \& De R\'ujula  2009b), where the 
ratio of binary neutron stars to ordinary stars is much higher than in the 
regular interstellar medium of galaxies. ICS of ambient light in GCs by the
highly relativistic jets, which produce the SHBs, can  explain the
origin of their extended emission (Dado, Dar \& De R\'ujula 2009b; 
Dado \& Dar 2017). SHB170817A did not take place in a GC or a bright 
location in its host galaxy NGC4993 (Levan et al.~2017) and, indeed,
as expected, no extended emission following its prompt emission was 
observed.

\section{The fireball afterglow}
In this chapter we show that the observed UVOIR afterglow of 
SHB170817A in the first two weeks after burst can be well explained by 
the expansion of an Arnett-type fireball (Arnett 1982) powered by 
several sources such as neutrino-anti neutrino annihilation outside 
the merging n*s (Goodman, Dar \& nussinov 1987), decay of radioactive 
elements within merger ejecta (Mactonova, Li \& Paczynski 1998) and by 
the radiation, high energy particles and relativistic winds emitted 
from the binary n*n* before the merger and the nascent n* 
after the merger.

It has been shown that the bolometric light curves of ordinary (Dado 
\& Dar 2013) and superluminous (Dado \& Dar 2015) SNe of Type Ia can 
be successfully described by a ``master formula" derived from 
Arnett-type models.  The underlying physics is complex, but the 
fitted values of the physical parameters turn out to be, in the cases 
that were studied, very close to the values expected and determined by 
simple physics considerations. To make this note self-contained we 
give in this section a brief description of the derivation of the 
master formula, based on energy conservation in the rest frame of 
a fireball powered by such energy source and loosing energy by expansion 
and radiation.

Let $t$ be the time after the beginning of the formation of a fireball. 
As long as it is highly opaque to optical photons and $\gamma$ rays, 
its thermal energy density is dominated by black body radiation,
${u(T)\!\approx\!7.56\times 10^{-15}\,T^4}$ ${\rm erg\, cm^{-3}\,K^{-4}}$, 
at a temperature $T$ that we assume for simplicity to be spatially 
uniform. The fireball's total radiation energy is $ U\!=\!V\,u$, with $V$ 
the fireball's volume. For a constant velocity of expansion, 
$ dV/dt\!=\!3\,V/t$ and the resulting energy loss is simply $U/t$, since
$ dU/dt\!=\!-p\,dV/dt$ and $ p\!=\!u/3$ for black-body radiation.

Photon emission constitutes a second mechanism of energy loss by
a fireball, corresponding to a bolometric luminosity $L\!\approx\! U/t_{d}$,
where the photon's mean diffusion time is $t_{d}\!\approx\! R^2/(c\,\lambda)$.
The photon's mean free path, $\lambda$, is dominated by their Thomson 
scattering on free electrons and positrons, 
$\lambda\!=\!1/(n_e\,\sigma_{_T})$, with $n_e\!\propto\! 1/R^3$ their number 
density. For a fireball expanding at a constant velocity, 
$R\!=\!v\,t$, whose total number of free electrons and positrons is
$N_e$, $t_d\!=\!t_r^2/t$, with $t_r^2\!=\!3\,N_e\,\sigma_{_T}/8\,\pi\,c\,v$.
For a Type Ia SN, $t_r$ can be estimated to be $\sim 11$ days 
(Dado \& Dar 2015).

Neutrino-antineutrino annihilation (Goodman, Dar \& Nussinov 1987), the decay of radioactive 
isotopes synthesized in the merger ejecta, (Lattimer \& Schramm 1974), magnetic dipole 
radiation, relativistic winds, and high energy particles emission from the n*s
contribute to  the energy balance within a fireball, at a rate $\dot E$. Gathering 
all three contributions to $\dot U$, the time variation of $U$, we conclude
that energy conservation implies :
\begin{equation}
\dot U\approx \dot E\!-\!U\,\left[{1\over t}\!+\!{t\over t_r^2}\right]
\label{Eq7}
\end{equation}
during the photospheric phase.

The solution of Eq.~(7) is
\begin{equation}
{ U\!=\!{e^{-t^2/(2\, t_r^2)}\over t}\,
\int_0^t \bar t\,e^{\bar t^2/(2\, t_r^2)} \dot{E}(\bar t)\,d\bar t}.
\label{Eq8}
\end{equation}

Consequently, the bolometric luminosity, $ L_b\!=\!t\,U/t_r^2$, is 
given by the simple analytic expression 
\begin{equation} { 
L_b\!=\!{e^{-t^2/(2\, t_r^2)}\over t_r^2} \int_0^t \bar t\,e^{\bar 
    t^2/(2\, t_r^2)} \dot{E}(\bar t)\, d\bar t}. 
 \label{Eq9} 
\end{equation}

This simple master formula, which was first derived by Dado \& 
Dar (2013), provides an excellent description of the bolometric 
light curve of Supernovae Type Ia and of superluminal supernovae 
(Dado \& Dar 2015).  For a short energy deposition time $t_d\!\ll 
\!t_r$ by neutrino annihilation and r-processes, the late-time 
($t\!>\!t_d $) behavior of Eq.(9) is

\begin{equation}
{ L_b\approx L(t_d)\,e^{-t^2/(2\, t_r^2)}}.
\label{Eq10} 
\end{equation}

Such a bolometric light-curve is expected if the compact remnant of 
the NSM170817 is a stellar black hole. A best fit of Eq.(10) to 
the bolometric light curve of SHB170817A reported by Smartt et 
al.~(2017), Evans et al.~(2017) and Pian et al.~(2017),  
shown in Figure 6, yields a rather unsatisfactory ${\chi^2/{\rm dof}=3.85}$
for ${t_r\!=\!3.94}$ d 
and ${L(t_d)\!=\!3.96\times 10^{41}}$ erg/s.

\begin{figure}[]
\centering
\epsfig{file=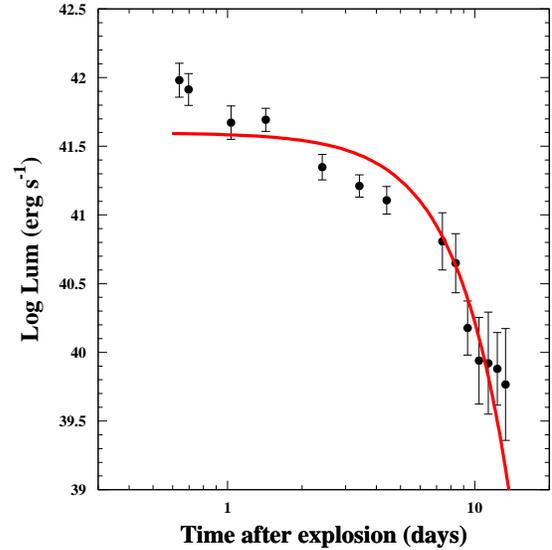,width=8.cm,height=8.cm}                                              
\caption{The best fit CB model bolometric light curve
of SHB170817A  to that reported by Smartt et al.~(2017), 
Evans et al.~(2017), and Pian et al.~(2017), assuming the 
compact remnant was a black hole.}                                                      
\label{Fig6}                  
\end{figure}      

For a fireball that at late time ($t>t_d$) is mainly powered by a pulsar 
with a period $P$ (Dar \& Dado 2017 and references therein) 
\begin{equation}
\dot{E}_{msp}(t)\!=\!L_{msp}(0) /(1\!+\!t/t_b)^2,
\label{Eq11}
\end{equation}
where $t_b\!=\!P(0)/2\dot{P}(0)$. 
The time dependence of $\dot{E}_{msp}$ is slow relative to that of the rest 
of the integrand in Eq.(9). As a consequence it is a good approximation
to factor it out of the integral, to obtain:
\begin{equation}
{L_b\!\approx\! L_{msp}(0)[1\!-\!e^{-t^2/2\, t_r^2}]/(1\!+\!t/t_b)^2}.
\label{Eq12}  
\end{equation}
A best fit of Eq.(12) to the  bolometric light
curve of SHB170817A reported by Drout et al.~(2017),
shown in Figure 7, 
yields ${L_{msp}(0)\!=\!2.27\times  10^{42}}$ erg/s,  ${ t_b\!=\!1.15}$ d,
and ${t_r\!=\!0.23}$ d,  with an entirely satisfactory $\chi^2/{\rm dof}\!=\!1.04$.
For these parameters, the approximation of Eq.(12)
differs from the ``exact" result of substituting Eq.(11) into
Eq.(9) by 15\% at the peak luminosity and $\!<\! 2$\% at $t\!>\!2$ d.

\begin{figure}[]
\centering
\epsfig{file=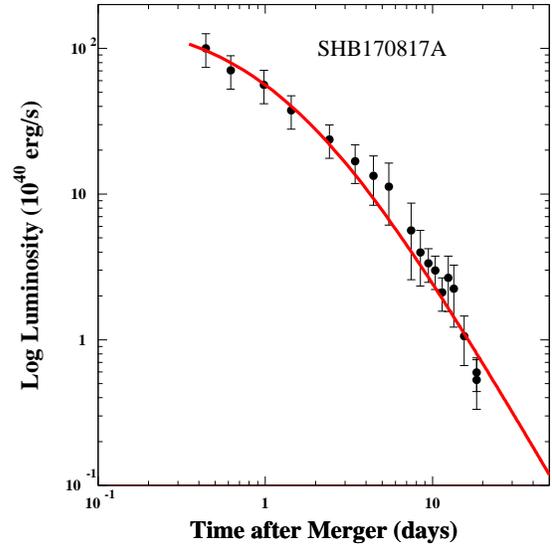,width=8.cm,height=8.cm}
\caption{The best fit CB model bolometric lightcurve
of SHB170817A  to that reported by Drout et al.~(2017),
assuming a neutron star remnant.}
\label{fig7}
\end{figure}

The CB model best fits to the measured bolometric light curves of the 
UVOIR afterglow of SHB170817A, as shown in 
Figures 6,7  suggest that the compact remnant 
of NSM170817 is probably a pulsar rather than a stellar mass black 
hole.

On the other hand, best fits to the bolometric light curve 
reported by Cowperthwaite et al.~(2017) and shown in 
Figures 8,9 are much less conclusive: 
${\chi^2/{\rm dof}\!=\!1.26}$ 
for ${t_r\!=\!3.70}$ d and ${L(t_d)\!=\!2.95\times 10^{41}}$ erg/s 
for a black hole remnant, while for a neutron star remnant,
${\chi^2/{\rm dof}\!=\!1.0}$  for
${\L_{msp}(0)\!=\!2.21\times 10^{47}}$ erg/s,   ${t_b\!=\!1.00}$ d,
and ${\rm  t_r\!\ll\! 1}$ d.

\begin{figure}[]
\centering
\epsfig{file=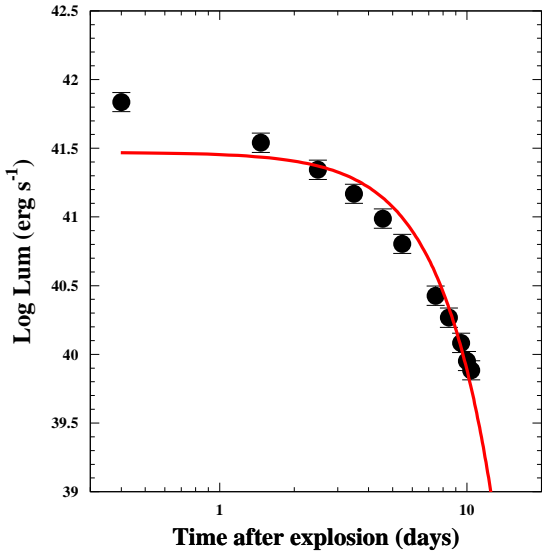,width=8.cm,height=8.cm}
\caption{The best fit CB model bolometric lightcurve 
of SHB170817A  to that reported in Cowperthwaite et al.~(2017),
assuming  a black hole  remnant.} 
\label{fig8}
\end{figure}

\begin{figure}[]
\centering
\epsfig{file=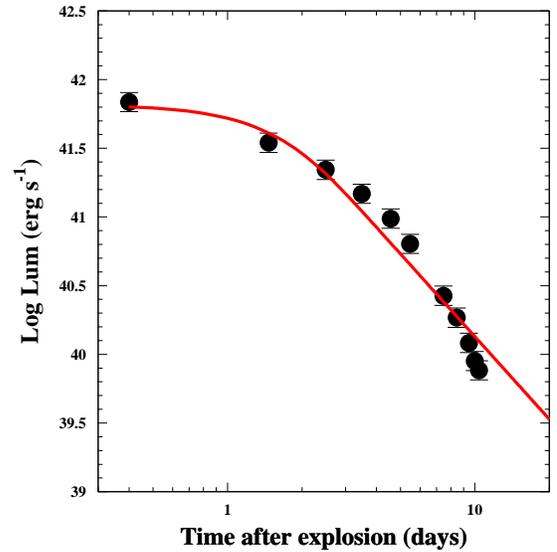,width=8.cm,height=8.cm}
\caption{The best fit CB model bolometric lightcurve 
of SHB170817A  to that reported in Cowperthwaite et al.~(2017),
assuming  a neutron star remnant.}
\label{fig9}
\end{figure}

\section{The Far Off-Axis Afterglow} 
The circumburst medium in the path of a CB moving with a Lorentz factor 
$\gamma\gg 1$ is completely ionized by the CB's radiation. The ions of the 
medium that the CB sweeps in generate within it turbulent magnetic fields.  
The electrons that enter the CB with a Lorentz factor $\gamma(t)$ in its 
rest frame are Fermi accelerated there, and cool by emission of synchrotron 
radiation, an isotropic afterglow in the CB's rest frame. As for the rest of 
the CB's radiations, the emitted photons are beamed into a narrow cone along 
the CB's direction of motion, their arrival times are aberrated, and their 
energies boosted by the Doppler factor $\delta(t)$ and redshifted by the cosmic 
expansion.

The observed spectral energy density of the {\it unabsorbed} synchrotron 
afterglow has the form (e.g., Eq.~(28) in Dado, Dar \& De R\'ujula 2009a)
\begin{equation}
F_{\nu}\!\propto\![\gamma(t)]^{3\,\beta\!-\!1}\,
[\delta(t)]^{\beta\!+\!3}\, \nu^{\!-\!\beta}\,,
\label{Eq13}
\end{equation}
where $\beta$ is the spectral index of the emitted radiation at a
frequency $\nu$. 

The swept-in ionized material decelerates the CB's motion. Energy-momentum 
conservation for such a plastic collision between a CB of baryon number 
$N_{_{\!B}}$, radius $R_{CB}$, and an initial Lorentz factor $\gamma_0\gg 1$ 
yields the deceleration law (e.g. Eq.~(3) in Dado and Dar 2012)
\begin{equation}
\gamma(t)\! =\!{\gamma_0\over [\sqrt{(1\!+\!\theta^2\,\gamma_0^2)^2\!+\!t/t_s}
         \!-\!\theta^2\,\gamma_0^2]^{1/2}}\,,
\label{Eq14}
\end{equation}
where $t_s\!=\!(1\!+\!z)\, N_{_{\!B}}/( 8\,c\, n\,\pi\, 
R_{CB}^2\,\gamma_0^3)$ 
is the slow-down time scale. 
The frequency and time dependence of the afterglow implied by 
Eqs.~(13),(14) depend only on  three parameters: 
the product $\gamma_0\,\theta$,  
the spectral index $\beta$, and the slow-down time-scale $t_s$. 
For  $t\!\gg\!t_b$
Eq.~(14) yields  $\gamma(t)\!\propto\! t^{\!-\!1/4}$. 
and consequently a power-law decline,
$F_\nu(t)\propto t^{\!-\!(\beta\!+\!1/2)}\,\nu^{\!-\!\beta}$.
independent of the values of $t_b$ and $\gamma(0)\, \theta$.   
For far off-axis GRBs and SHBs, 
as long as $\gamma^2\,\theta^2\!\gg\!1$,  
$F_\nu(t)$ for $t<t_b$, rises like   
\begin{equation}
F_\nu(t)\propto t^{1\!-\!\beta/2}\,\nu^{\!-\!\beta}\,.
\label{Eq15}
\end{equation}
In Figure 10 we compare the 6 GHz light curve of the radio afterglow of 
SHB170817A first discovered by Hallinan et al. (2017) and followed up with 
the Karl G. Jansky Very Large Array (VLA), the Australia Telescope Compact 
Array (ATCA) and the upgraded Giant Metrewave Radio Telescope (uGMRT) and 
summarized by Mooley et al.(2018), and the CB model prediction. The CB 
model light curve was obtained from Eq.(13) with the  measured radio 
spectral index $\beta\!=\!0.57\!\pm\! 0.09$ and Eq. (14) for 
$\gamma(t)$ where the value $\gamma(0)\,\theta\!=\!5$ was obtained 
from the measured value $E_p\!=\!82$ keV and the deceleration time scale 
$t_s\!=\!0.167$ days from a best fit to the data.

\begin{figure}[]
\centering
\epsfig{file=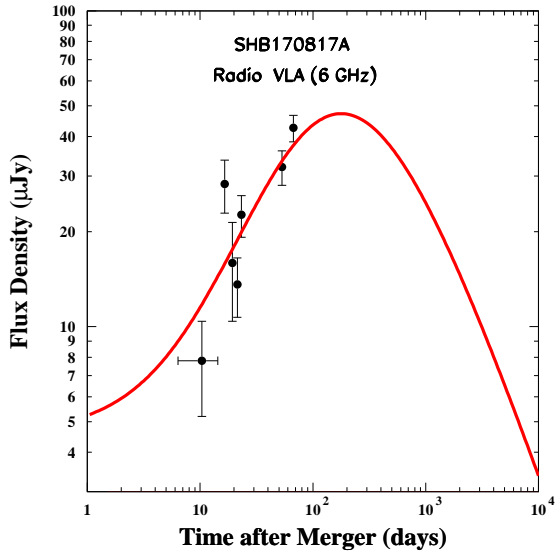,width=8.cm,height=8.cm}
\caption{Comparison between the 6 GHz light curve of the afterglow of 
SHB170817A  measured with the VLA (Hallinan et al. 2017, Mooley et al. 2018)
and the light curve expected in the CB model as described in the text.}
\label{fig10}
\end{figure}

In Figure 11 we compare the X-ray light curve measured with the 
Chandra X-ray observatory (CXO) and that predicted by Eqs.(13), and (14) 
using the same parameters as in Figure 10.  Either 
Eq.(13) with Eq.(14), or Eq.(15) describe well the  rising 
phase of the $0.3\!-\!10$ keV X-ray light curve measured with CXO 
(Troja et al. 2017a,b; Margutti et al. 2017a,b; Haggard et al. 2017).
The best fit value $t_s\!=\!0.167$ d yield the product 
$n_b\,R_{CB}^2\!=\!2.3\times 10^{22}$/cm, but not the individual values 
of the baryon density $n_b$ of the circumburst interstellar density 
and of $R_{CB}$, the rdius of the CB.

\begin{figure}[]
\centering
\epsfig{file=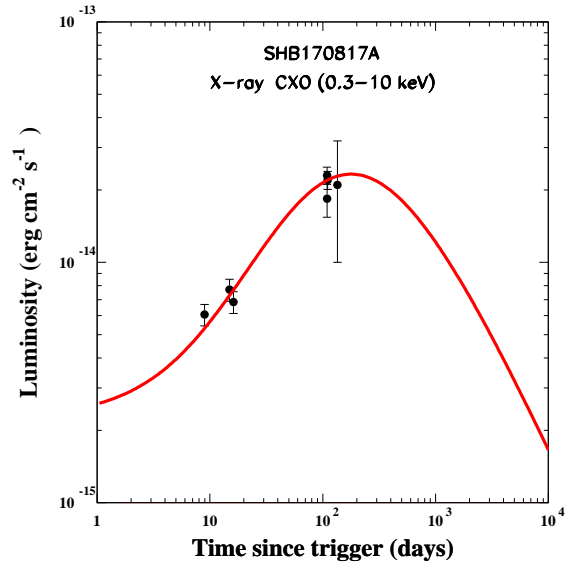,width=8.cm,height=8.cm}
\caption{Comparison between the light curve of the X-ray afterglow of
SHB170817A  measured with the CXO (Troja et al.                      
2017a,b; Margutti et al. 2017a,b; Haggard et al. 2017)
and the light curve expected in the CB model as described in the text.}
\label{fig11}
\end{figure}

Further optical and FIR observations of the counterpart of GW170817 with 
the Hubble Space Telescope, which took place on 6 Dec 2017 (Levan et al. 
2017) recovered the source in optical filters, but did not detect it in the 
infrared, where the background from the galaxy is higher. The measured 
magnitudes of the source in the optical bands are broadly consistent with 
the extrapolation from the 93 day radio epoch (Mooley et al. 2018)  to the 
near contemporaneous observations with CXO (Troja et al. 2017a,b; Margutti et 
al. 2017a,b; Haggard et al. 2017).

\section{Superluminal motion}
\label{sec:SL}

A very specific prediction of the CB model concerns the apparently 
superluminal motion in the plane of the sky, relative to the engine that 
produced them (Dar \& De R\'ujula 2000b,; Dado, Dar \& De R\'ujula 2016 and 
references therein), of the CBs moving towards the observer at a small but 
not vanishing angle $\theta$.
The apparent sky velocity of a CB relative to its emission point is given by   
\begin{equation}
V_{\rm app}\!\approx\!  [2\,\gamma(t)^2\,\theta/(1\!+\!\theta^2\gamma(t)^2)]\,c
\label{Eq16}
\end{equation}
with $\gamma(t)$ as in Eq.(14). For  $\gamma(0)\,\theta \!\approx\! 5$  as estimated 
in Section 2  and $t_s\!=\!0.167$ days as estimated in Section 6, $V_{\rm app}\!\sim\! 
2\,c/\theta\approx 400\,c$ for $t\!\ll\! 120$ d, and
$V_{\rm app}\!\approx\! 60\,c\, (t/10^3\,{\rm d})^{-1/2}$ for 
$t\!\gg\!120$ d. The estimated superluminal speed of the CB as function of time 
as given by Eq(16) is shown in Figure 12.

\begin{figure}[]
\centering
\epsfig{file=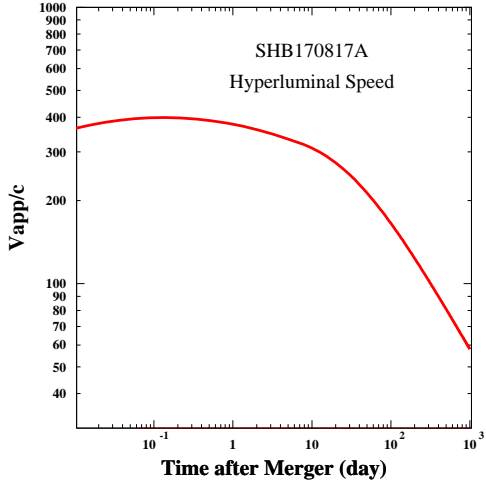,width=7.cm,height=7.cm}
\caption{The estimated superluminal speed of the CB  relative to the location 
of SHB170817A as a function of time after burst.}
\label{fig12}
\end{figure}

The angular displacement $\alpha(t)$ from the location of the 
neutron star merger to the CB's later position is:
\begin{equation}
\alpha(t)\!=\!c\,\int_0^t dt'\,V_{\rm app}(t')/D_A\, ,
\label{Eq17}
\end{equation}
with $D_A\!=\!39.6$ Mpc the angular distance to SHB170817A in the standard 
cosmology. 

In a VLA or VLBI observation of SHB170817A, the angular Fresnel 
scale $\sqrt{\lambda/(2\,\pi\,D_A)}$ is of order $0.1\,\mu$as, 
considerably smaller that the angular size of a CB. This may lead 
one to expect that a CB's image would scintillate (Taylor et 
al.~2004). But typical integration times of these observations are 
100 minutes. At an early time of observation ($t\!\sim\! 0$), the image 
of the CB of SHB170817A would move by an angle $117\,\mu$as, and 
$48\,\mu$as at day $t\!=\!300$. This shifting position while the data 
are accumulated would obliterate the scintillations (Dado et al. 2016).

\section{Conclusions}

The temporal and spatial coincidence of NSM170817 and SHB170817A has 
shown that at least a fraction, if not most SHBs, are produced in NSMs.

SHB170817A and its afterglow at redshift $z=0.00936$ ($D_L\approx 40$ 
Mpc) appeared to be very different from all other SHBs with known 
redshift, including SHB130603B at redshift $z=0.3564$ ($D_L\approx 2000$ 
Mpc), where a faint infrared emission was claimed to be the first 
observational evidence for a macronova produced by a NSM (e.g., Berger et 
al. 2013, Tanvir et al. 2013). But the observations of the low luminosity 
SHB170817A and its afterglow produced by NSM170817 can be naturally 
explained, as we have shown, by the cannonball model of GRBs, if 
SHB170817A was beamed along a direction far off from its line of sight 
and took place in the low density and low luminosity  of its host 
galaxy NGC 4993, rather than in a dense stellar region.
The CB model analysis of SHB170817A implies that 
like SN1998bw-GRB980425, GW170817-SHB170817A probably was a very rare 
event.

In fact, based on the cannonball model of SHBs, we predicted (Dado 
\& Dar 2017) before GW170817 that Ligo-Virgo detections of NSMs 
will be followed if at all, only by low-luminosity (far off-axis) 
SHBs, or by orphan SHB afterglows but not by ordinary SHBs. This is 
due to the much smaller detection horizon of Ligo-Virgo compared to 
the mean distance of SHBs estimated from SHBs with known redshift.

Our detailed analysis indicates that the compact remnant of NSM170817 
probably was a neutron star and not a black hole, in agreement with the 
evidence from the afterglow of all other SHBs with known redshift and 
well sampled X-ray afterglow (Dado \& Dar 2017).
  
The existence of a glory with a radius $\sim 10^{15}$ cm around the merger 
time is required by the CB model analysis of SHB170817A. Its origin is not 
clear. The most likely explanation is a PWN surrounding the binary neutron 
stars and powered by the emission of radiation and high energy particles 
and winds by one or both of them LONG before the merger. A fast expanding 
macronova powered by the radioactive decay of r-processed elements which 
exists already a day before the NSM when the merging neutron stars were 
separated by more than 100 times when they collided does not seem plausible. 
Observations of more NSM-SHB events and theoretical studies will be required 
in order to identify the origin of the UVOIR fireball.

We have shown that, in spite of the CB-model's simplicity -the description 
of the $\gamma$-ray pulse by ICS and of the late afterglow by synchrotron 
radiation from a superluminally moving CB-  the underlying model intrinsic and 
environmental parametres we extracted from both the pulse and the afterglow 
are very consistent. Moreover, they are compatible with our previous work on
GRB and SHB pulses and afterglows.

The large superluminal speed and angular displacement that we have 
discussed could perhaps be observed by the VLA and VLBI follow-up 
measurements of the late-time location of the radio afterglow of the 
jet, relative to the location of SHB170817A. Such observations would 
be most decisive if a non-vanishing angular displacement could be 
measured at least two different times. It goes without saying that the 
precise values of the predicted displacement are quite uncertain, the 
crucial point being the trend shown in Figure 12.

{\bf Acknowledgments:} We are particularly grateful to a referee who
pointed out that, to be more complete, we ought to extract 
the parameters describing the cannonball and its surroundings.\\
ADR acknowledges that this project has received 
funding/support from the European Union Horizon 2020 research and 
innovation programme under the Marie Sklodowska-Curie grant agreement No 
690575.

\end{document}